\documentclass[review,preprint,10pt,3p]{elsarticle}
\usepackage{graphicx}
\usepackage[colorlinks = true,
            linkcolor = cyan,
            urlcolor  = blue,
            citecolor = red,
            anchorcolor = blue]{hyperref}\usepackage{color}
\usepackage{amssymb}
\usepackage{multirow}
\usepackage{amsthm}
\usepackage{textcomp}
\usepackage{mathtools}
\usepackage{lineno}
\journal{Nuclear Instruments and Methods in Physics Research Section A}

\usepackage{comment}
\usepackage[separate-uncertainty,retain-explicit-plus,per-mode = symbol]{siunitx}
\usepackage{url}

\newcommand{\bet}{\mbox{$\beta$}}
\newcommand{\alp}{\mbox{$\alpha$}}
\newcommand{\gr}{\mbox{$\gamma$-ray}}
\newcommand{\grs}{\mbox{$\gamma$-rays}}
\newcommand{\E}[1]{\mbox{$\times10^{#1}$}}
\newcommand{\hydrogen}{\mbox{$^{1}$}H}

\newcommand{\paone}{\mbox{$^{231}$Pa}}
\newcommand{\pafourm}{\mbox{$^{234m}$Pa}}

\newcommand{\bifour}{\mbox{$^{214}$Bi}}

\newcommand{\pbten}{\mbox{$^{210}$Pb}}
\newcommand{\pbnine}{\mbox{$^{209}$Pb}}

\newcommand{\rntwo}{\mbox{$^{222}$Rn}}
\newcommand{\tho}{\mbox{$^{232}$Th}}
\newcommand{\thothirty}{\mbox{$^{230}$Th}}

\newcommand{\ura}{\mbox{$^{238}$U}}
\newcommand{\urafour}{\mbox{$^{234}$U}}

\newcommand{\rasix}{\mbox{$^{226}$Ra}}

\newcommand{\urafive}{\mbox{$^{235}$U}}
\newcommand{\tlten}{\mbox{$^{210}$Tl}}

\newcommand{\alphan}{\mbox{$(\alpha,\MakeLowercase{n})$}}
\newcommand{\jendl}{\mbox{JENDL}}
\newcommand{\empire}{\mbox{EMPIRE-2.19}}
\newcommand{\sources}{\mbox{SOURCES-4C}}
\newcommand{\srim}{\mbox{SRIM}}
\newcommand{\talys}{\mbox{TALYS}}
\newcommand{\nucbot}{\mbox{NeuCBOT}}

\newcommand{\gerda}{\mbox{GERDA}}
\newcommand{\exo}{\mbox{EXO}}

\newcommand{\website}{\mbox{\url{https://github.com/shawest/neucbot}}}

\begin{document}

\title{Radiogenic Neutron Yield Calculations for Low-Background Experiments}

\author[add]{S. Westerdale\corref{cor1}\fnref{fn1}}
\ead{shawest@physics.carleton.ca}
\cortext[cor1]{Corresponding author}
\fntext[fn1]{Current address: Department of Physics, Carleton University, Ottawa, Ontario K1S 5B6, Canada}
\author[add]{P. D. Meyers}
\address[add]{Department of Physics, Princeton University, Princeton, New Jersey 08544, USA}

\date{\today}

\begin{abstract}

Nuclear recoil backgrounds are one of the most dangerous backgrounds for many dark matter experiments. A primary source of nuclear recoils is radiogenic neutrons produced in the detector material itself. These neutrons result from fission and \alphan\ reactions originating from uranium and thorium contamination. 
In this paper, we discuss neutron yields from these sources. We compile a list of \alphan\ yields for many materials common in low-background detectors, calculated using \nucbot\ (Neutron Calculator Based On \talys), a new tool introduced in this paper, available at \website. These calculations are compared to computations made using data compilations and \sources.
\end{abstract}

\begin{keyword}
 dark matter \sep neutron backgrounds \sep \alphan\ yields \sep \nucbot
\end{keyword}

\maketitle



\section{Introduction}
Neutron backgrounds are often considered among the dominant backgrounds in many low-background experiments, such as neutrino-less double $\beta$-decay searches and direct dark matter detectors. 

Neutrino-less double \bet-decay experiments generally rely on having a high energy resolution and very little background near the endpoint of the double \bet-decay spectrum. While experiments generally take great paints to avoid having radioactive contaminants that produce a signal in the relevant energy range, such backgrounds may be produced by neutron interactions. Fast neutrons may produce a signal in the region of interest as they thermalize, or they may produce \grs\ that can produce such signals when they capture on detector materials. For example, \exo\ includes signals produced by neutron captures on $^{136}$Xe, \hydrogen, $^{65}$Cu, and $^{63}$Cu among their dominant backgrounds~\cite{exo-200_collaboration_investigation_2015}. Neutron-induced backgrounds such as the  $^{76}\text{Ge}(n,\gamma)^{77}\text{Ge}$ reaction are considered important potential backgrounds in \gerda~\cite{ackermann_gerda_2013}, and neutron captures on \hydrogen\ are expected to be among the primary backgrounds in SNO+~\cite{andringa_current_2016}.


In dark matter experiments looking for Weakly Interacting Massive Particles (WIMPs), electromagnetic backgrounds such as \grs\ and \bet-decays are expected to scatter off of electrons in the detector, while neutrons and nuclei are expected to scatter on target nuclei. Since WIMPs are expected to produce nuclear recoils, techniques that discriminate between electron and nuclear recoils are very effective at removing backgrounds. However, neutron-induced nuclear recoils remain an important background for these experiments, since they can produce a signal similar to what is expected from WIMPs~\cite{DarkSide_UAr_2015,Boulay_deap-3600_2012,aprile_lowering_2015,Malling_after_2011,CDMS2015_Ge}.


Nuclear recoil backgrounds result from \alp-decays on the inner surface of the detector ejecting a nucleus into the active volume of the detector or from neutrons scattering in the active volume.
Neutrons may be cosmogenic in origin if they are produced by cosmogenic muons interacting with the environment through processes such as spallation, or they may be radiogenic, primarily resulting from spontaneous fission of \tho, \ura, and \urafive, or from the \alphan\ reaction occurring when \alp-particles produced in these decay chains interact with nuclei in the material they are traveling in.

Cosmogenic neutrons are typically mitigated by moving a detector deep underground, where the muon flux is greatly reduced. External muon vetoes, as described in~\cite{agnes_veto_2016}, can be used to tag muons coincident with nuclear recoils produced by neutrons. A detailed discussion of these backgrounds is provided in~\cite{empl_fluka_2014}. The focus of this document will be radiogenic neutrons.

Radiogenic neutrons result from nuclear interactions within a given material. While they can be reduced with careful material selection, some amount will inevitably remain. These neutrons may scatter once in the sensitive volume of a detector and then leave, producing a signal identical to that expected from a WIMP. While external veto systems, such as the design discussed in~\cite{agnes_veto_2016,westerdale_prototype_2016}, may be able to tag these backgrounds, a quantitative description of the radiogenic neutron backgrounds in a detector is necessary for any low-background experiment to understand and minimize its backgrounds.

\section{Decay Chains and Secular Equilibrium}\label{sec:decaychains}
In this discussion, we will focus on three decay chains: \tho, \ura, and \urafive. Isotopes in these chains are expected to produce most of the radiogenic neutron backgrounds present in most low-background experiments. Typically, these experiments strive to reduce their contamination of these isotopes through screening campaigns, such as \gr\ spectroscopy, as discussed in~\cite{maneschg_measurements_2008}, glow-discharge mass spectrometry, as discussed in~\cite{gdms_evans}, and other techniques.

We will assume secular equilibrium in these chains, with a few exceptions. 

Secular equilibrium may be broken where a long-lived gaseous isotope can emanate from a material, where manufacturing or purification processes may selectively remove some isotopes, where the material may have been exposed to elements partway down a decay chain, such as may be the case for radon, or where other chemical proceses may differently affect elements with different chemical properties such as solubility in water. If these isotopes or their precursors have half-lives longer than the scale of the experiment, it may not be appropriate to assume secular equilibrium for the entire decay chain. In these cases, we will divide the chains into sub-chains in which we expect secular equilibrium to be preserved with respect to the top of the sub-chain.

In particular, \ura\ has a half-life of 4.5\E{9}\,years and decays to \rasix, with a half-life of 1600\,years, through four intermediate isotopes.
Since \pafourm, one of the precursors of \rasix, emits a 1\,MeV \gr\ that can be measured experimentally and \rasix\ has a 186\,keV \gr, we split the \ura\ decay chain into an upper and lower chain, defining all isotopes before \rasix\ to be in the upper chain, and \rasix\ and its progeny to be in the lower chain. 
One possible explanation for the chain breaking at \rasix\ is that radium is an alkaline earth metal, while its precursors are all actinides. Since these two groups have different chemical properties, it is possible that some chemical processes can affect these elements differently. Since \rasix\ has a very long half life, secular equilibrium will remain broken if \rasix\ is taken out of equilibrium with its precursors.
Furthermore, since \pbten\ has a half-life of 22.2\,years, far longer than its precursors, excess \pbten\ may accumulate in materials due to \rntwo\ exposure in the air, causing secular equilibrium to be broken once again. 
Therefore, we account for the \pbten\ decay chain, consisting of \pbten\ and its progeny. 

Throughout this document, we will discuss neutron yields in terms of neutrons produced per decay of the top of the decay chain. For a decay chain in secular equilibrium, this includes neutrons produced by all of the isotopes in this chain, weighted by the relevant branching ratios.

\section{Direct Neutron Emission and Spontaneous Fission}
While the primary focus of this document will be on neutrons produced by the \alphan\ process, it is worth drawing attention to two other processes that produce radiogenic neutrons: direct neutron emission and spontaneous fission. The rates of both of these processes depend only on the amount of uranium and thorium present in each detector component, and not on the material in which the contamination is present. In materials with particularly low \alphan\ cross sections, these other processes may contribute significantly to the neutron background. We therefore discuss these radiogenic neutron sources for comparison.

\bifour\ \alp-decays to \tlten\ with a branching ratio of 0.021\%, which then \bet-decays to \pbten. There is a 0.007\% chance that this \bet-decay will go to an excited state of \pbten, which decays by emitting a 200--260\,keV neutron to \pbnine~\cite{basunia2014nuclear}. Due to these branching ratios, we expect to see these neutrons in $\sim$1.5\E{-8} of all decays of the lower \ura\ chain.

Heavy nuclei that ordinarily \alp-decay may instead fission into smaller nuclei. When this happens, many particles may be ejected as well, including several MeV-scale \grs\ and some number of neutrons. The distribution of the number of \grs\ produced is discussed in~\cite{valentine_evaluation_2001}, and the distribution of the number of neutrons emitted is discussed in~\cite{holden1984reevaluation}. These studies showed that the number of neutrons emitted in the spontaneous fission of \ura\ can be modeled by a Gaussian distribution with a cutoff at 0, a mean of 2.05$\pm$0.04, and a standard deviation of 1.04$\pm$0.03. 

These fission reactions are described in Table~\ref{tab:fissionns}, which summarizes the spontaneous fission branching ratio BR$_{\text{SF}}$, the mean neutron kinetic energy $\langle E\rangle$, the mean neutron multiplicity $\langle\nu\rangle$, and the total number of neutrons produced per second per Becquerel of each decay chain for each isotope in these chains that may undergo spontaneous fission. 

\begin{table}[ht]
 \centering
 \caption{Spontaneous fission branching ratios, mean neutron energies in MeV, mean neutron multiplicities, and neutron yields in n/s/Bq, calculated using \sources.}
 \begin{tabular}{ll|cccc}\hline\hline
  Chain                            & Iso.                & BR$_{\text{SF}}$ & $\langle E\rangle$ & $\langle\nu\rangle$ & Yield       \\\hline
  \multirow{1}{*}{\small \tho}     & {\small \tho}       & 1.80E-11         & 1.60               & 2.14                & 3.85E-11    \\\hline
  \multirow{3}{*}{\small \ura}     & {\small \ura}       & 5.45E-07         & 1.69               & 2.01                & 1.10E-06    \\
                                   & {\small \urafour}   & 1.64E-11         & 1.89               & 1.81                & 2.97E-11    \\
                                   & {\small \thothirty} & 3.8E-14          & 1.71               & 2.14                & 8.13E-14    \\\hline
  \multirow{2}{*}{\small \urafive} & {\small \urafive}   & 1.60E-03         & 1.89               & 1.93                & 3.09E-13    \\
                                   & {\small \paone}     & 7.00E-11         & 1.93               & 1.86                & 1.30E-10     \\\hline
  \hline
 \end{tabular}
 \label{tab:fissionns}
\end{table}

\section{$(\alpha,n)$ Neutrons}
The \alphan\ reaction occurs predominantly in low-to-mid-$Z$ materials with contamination from \alp-emitting isotopes. When these isotopes decay, the emitted \alp\ particle may capture on another nucleus in the material to form a compound nucleus, which may decay by neutron emission. For the calculations discussed here, we consider a thick target in which the \alp\ particle captures in the same material in which it was produced. Calculations of the neutron yield from the \alphan\ reaction (\emph{i.e.}, the \alphan\ yield) therefore depend on the energy spectra of \alp-decays and the elemental and isotopic composition of the material. These calculations also depend on the stopping power of \alp\ particles of a given energy in the material as well as the \alphan\ cross sections and the structure of nuclei involved in these reactions.

While these neutrons are sometimes accompanied by a \gr, either correlated with the decay of the \alp-emitter or from the relaxation of the final nucleus, neutrons are also often produced alone. This possibility may make \alphan\ neutrons particularly troublesome backgrounds, as there may be no accompanying signal to help tag the neutron. The rest of this document will therefore be focused on calculating \alphan\ yields, including the introduction of \nucbot\ as a tool for calculating these yields (see Section~\ref{subsec:nucbotcalcs}). In order to benchmark \nucbot\ against other standards, we will calculate \alphan\ yields for several materials using \nucbot\ and compare these yields to calculations performed using measured yields on individual isotopes (see Section~\ref{subsec:measuredyieldcalcs}) as well as yields and neutron energy spectra predicted by \sources\ (see Section~\ref{subsec:sourcescalcs}).


\subsection{Materials Considered}

\begin{table}
 \centering
 \caption{Material compositions used in \alphan\ calculations.}
 \footnotesize
 \begin{tabular}{l|ccccc|ccccc}\hline\hline
  Material                                               & \multicolumn{5}{c|}{Composition (\% mass)}     & \multicolumn{5}{c}{Composition (\% element)}     \\\hline
  \multirow{2}{*}{Acrylic}                               & H    & C    & O    &      &                    & H    & C     & O                                 \\
                                                         & 8.1  & 60.0 & 31.9 &      &                    & 53.3 & 33.3  & 13.3                              \\\hline
  \multirow{2}{*}{Alumina}                               & Al   & O    &      &      &                    & Al   & O                                         \\
                                                         & 52.9 & 47.1 &      &      &                    & 40.0 & 60.0                                      \\\hline
  \multirow{2}{*}{Aluminum}                              & Al   &      &      &      &                    & Al                                               \\
                                                         & 100  &      &      &      &                    & 100                                              \\\hline
  \multirow{2}{*}{Argon}                                 & Ar   &      &      &      &                    & Ar                                               \\
                                                         & 100  &      &      &      &                    & 100                                              \\\hline
  \multirow{2}{*}{Be-Cu Alloy}                           & Be   & Ni   & Cu   &      &                    & Be   & Ni   & Cu                                 \\
                                                         & 0.4  & 1.8  & 97.8 &      &                    & 2.7  & 1.9  & 95.4                               \\\hline
  \multirow{1}{*}{Borosilicate Glass}                    & \multicolumn{5}{c|}{---}                       & \multicolumn{5}{c}{---}                          \\\hline
  \multirow{2}{*}{Brass}                                 & Cu   & Zn   &      &      &                    & Cu   & Zn                                        \\
                                                         & 63   & 37   &      &      &                    & 63.7 & 36.3                                      \\\hline
  \multirow{2}{*}{Cirlex}                                & H    & C    & N    & O    &                    & H    & C    & N    & O                           \\
                                                         & 2.6  & 69.1 & 7.3  & 20.9 &                    & 25.4 & 56.6 & 5.1 & 12.9                         \\\hline
  \multirow{2}{*}{Copper}                                & Cu   &      &      &      &                    & Cu                                               \\
                                                         & 100  &      &      &      &                    & 100                                              \\\hline
  \multirow{2}{*}{Fused Silica}                          & Si   & O    &      &      &                    & Si   & O                                         \\
                                                         & 46.7 & 53.3 &      &      &                    & 33.3 & 66.7                                      \\\hline
  \multirow{6}{*}{Kovar}                                 & Fe   & Ni   & Co   & Mn   & Si                 & Fe   & Ni   & Co   & Mn   & Si                   \\
                                                         & 52.5 & 29   & 17   & 0.3  & 0.2                & 41.69& 21.92& 12.79& 0.24 & 0.32                 \\
                                                         & C    & Al   & Mg   & Zr   & Ti                 & C    & Al   & Mg   & Zr   & Ti                   \\
                                                         & 0.02 & 0.1  & 0.1  & 0.1  & 0.1                & 22.15& 0.16 & 0.18 & 0.05 & 0.09                 \\
                                                         & Cu   & Cr   & Mo   &      &                    & Cu   & Cr   & Mo                                 \\
                                                         & 0.2  & 0.2  & 0.2  &      &                    & 0.14 & 0.17 & 0.09                               \\\hline
  \multirow{1}{*}{NUV-HD Silicon}                        & Si   & O    & Al   & Ti   &                    & Si   & O    & Al   & Ti                          \\
  Photomultipliers                                       & 99.27& 0.5  & 0.2  & 0.03 &                    & 98.90& 0.87 & 0.21 & 0.02                        \\\hline
  \multirow{1}{*}{Polyethylene}                          & H    & C    & O    &      &                    & H    & C    & O                                  \\
  Terephthalate (PET)                                    & 4.2  & 62.5 & 33.3 &      &                    & 36.4 & 45.4 & 18.2                               \\\hline
  \multirow{2}{*}{PTFE}                                  & C    & F    &      &      &                    & C    & F                                         \\
                                                         & 24.0 & 76.0 &      &      &                    & 33.3 & 66.7                                      \\\hline 
  \multirow{2}{*}{Solder}                                & Sn   & Ag   & Cu   &      &                    & Sn   & Ag   & Cu                                 \\
                                                         & 96.5 & 3.0  & 0.5  &      &                    & 95.8 & 3.3  & 0.9                                \\\hline
  \multirow{4}{*}{Stainless Steel}                       & C    & Cr   & Mn   & Ni   & P                  & C    & Cr   & Mn   & Ni   & p                    \\
                                                         & 0.04 & 18   & 2    & 8    & 0.05               & 0.2  & 18.8 & 2.0  & 7.4  & 0.1                  \\
                                                         & S    & Si   & N    & Fe   &                    & S    & Si   & N    & Fe                          \\
                                                         & 0.03 & 1    & 0.1  & 70.9 &                    & 0.1  & 1.9  & 0.4  & 69.1                        \\\hline
  \multirow{2}{*}{Titanium}                              & Ti   &      &      &      &                    & Ti                                               \\
							 & 100  &      &      &      &                    & 100                                              \\\hline
  \multirow{2}{*}{Viton}                                 & H    & C    & F    &      &                    & H    & C    & F                                  \\
                                                         & 0.9  & 28.1 & 71.0 &      &                    & 13.3 & 33.3 & 53.4                               \\\hline
  \multirow{2}{*}{Xenon}                                 & Xe   &      &      &      &                    & Xe                                               \\
							 & 100  &      &      &      &                    & 100                                              \\\hline
  \hline
  \end{tabular}
 \label{tab:matcomp}
\end{table}

The \alphan\ reaction rate and neutron spectrum depend on both the energy of the \alp\ particle being emitted and the various nuclei with which the emitted \alp\ particle interacts. 
Since the \alphan\ reaction depends on the nuclei with which the emitted \alp\ particle may interact, either through their contribution to the stopping power or through the \alphan\ reaction itself, it is therefore important to define the chemical compositions of the materials for which we are calculating \alphan\ yields.

We summarize the chemical compositions used for these calculations in Table~\ref{tab:matcomp}. The same chemical and isotopic compositions were used for calculations performed using \nucbot, \sources, and measured yields. Notably, \sources\ requires that elemental and isotopic compositions be specified by the fraction of total atoms and nuclei that are a given element or isotope. The mass fractions given in Table~\ref{tab:matcomp} were therefore converted to isotopic fractions for \sources\ calculations. For each element, we assume natural isotopic abundances as reported in~\cite{de_bievre_table_1993}.

Since the same material compositions were assumed for calculations performed using \nucbot, measured yields, and \sources, uncertainties in material compositions do not affect the comparison between the three different methods. Nevertheless, in order to understand the uncertainties in the yields reported in this document, we discuss the uncertainties in these material compositions, based on typical tolerances reported in literature.


While there are multiple Be-Cu alloys, the composition used here is the one reported by Materion~\cite{materion_protherm_becualloy} for the PROtherm material.
Based on the uncertainties in the chemical composition reported by Materion, we estimate a $\sim47\%$ uncertainty in the total \alphan\ yield due to uncertainties in the chemical composition.

The composition of borosilicate glass was provided by Hamamatsu Photonics through private communications, and we were asked not to disseminate this information. While we cannot provide the actual composition here, the composition we assumed is that used by Hamamatsu in the stems (backplates with leads) of their R11065 photomultiplier tubes. 
The \alphan\ yield of borosilicate glass is dominated by boron, followed by lithium and aluminum. Based on the uncertainties in the chemical composition reported by Hamamatsu, we estimate a 5.7\% uncertainity in the total \alphan\ yield.

There are several different alloys of brass. For these calculations, we assumed a chemical composition typical of the ``common brass'' alloy. 
Based on typical tolerances reported in the composition of common brass, we estimate a 6.3\% uncertainty in the total \alphan\ yield due to uncertainties in the chemical composition.

Cirlex is an adhesiveless Kapton (polymide) laminate used in low-background circuit boards, and Viton is a fluoropolymer rubber that is commonly used to make o-rings. The chemical compositions used for these calculations are from the NIST ESTAR database~\cite{berger1998stopping}. We assume that the uncertainty in the chemical composition of these materials is small.

Kovar is a metal alloy designed to have the same thermal expansion coefficient as borosilicate glass, used in the outer shell of some Hamamatsu photomultiplier tubes. The composition used for these calculations is reported by the Carpenter Technology Corporation~\cite{carpenter_kovar_datasheet}. 
Based on the uncertainties in the chemical composition reported in~\cite{carpenter_kovar_datasheet}, we estimate a 0.01\% uncertainty in the total \alphan\ yield due to uncertainties in the chemical composition.

The chemical composition used for silicon photomultipliers in calculations was provided by Fondazione Bruno Kessler, for their NUV-SD devices. We assume that the uncertainty of the chemical composition of these devices is small.

The solder composition used for these calculations is based on the SAC305 ALPHA-Lo\textsuperscript{\textregistered} alloy produced by Pure Technology, which is used in the Indium3.2 lead-free solder produced by the Indium Corporation. 
Based on uncertainties in the chemical composition reported by Pure Technology, we estimate a 0.02\% uncertainty in the total \alphan\ yield due to uncertainties in the chemical composition.

For stainless steel, we used the nominal composition of 304L stainless steel reported by~\cite{ssa_stainless_specs}, assuming all elements whose compositions are given as upper bounds are at their limits. 
Based on the ranges provided for the concentrations of each constituent element, we estimate  3\% uncertainty in the \alphan\ yield due to uncertaintes in the chemical composition.

\subsection{Calculation using \nucbot}\label{subsec:nucbotcalcs}
To determine the \alphan\ yield of materials exposed to a given list of \alp\ particle energies or \alp-emitting nuclei, we have written a program that compiles output from \talys~\cite{koning2013talys} with nuclear decay information from the ENSDF database~\cite{tuli_evaluated_1996} and stopping power calculations from \srim~\cite{ziegler1985stopping}. 

We call this program \nucbot\ (Neutron Calculator Based On TALYS). It can be downloaded at \website.

\srim\ is a program written by Ziegler et al. for simulating ion propagation in materials, based on the work in~\cite{ziegler1985stopping}.

\talys\ is a general nuclear reaction simulation program that uses nuclear structure data and theoretical models to calculate nuclear reaction cross sections and emission spectra for a projectile particle at a given energy impinging upon a specific target nucleus. Validation of this code is discussed in~\cite{koning2013talys}. \talys\ uses the nuclear structure of the target, compound, and daughter nuclei to predict the cross sections for forming all of the energetically accessible excited states of the daughter nucleus, and the effects of the daughter nucleus's energy level are propagated into the outgoing neutron spectrum. These effects are propagated through the calculations performed by \nucbot\ using the \talys\ data.



Since \alphan\ yields may vary a lot between different target isotopes of a given element, we consider each target isotope in the material separately. If target isotope $i$ has a number density equal to $\eta_i$, the yield $Y_i(E_\alpha,E_n)$ of neutrons 
at energy $E_n$ by an \alp\ particle with energy $E_\alpha$ that travels a distance $dx$ is given by
\begin{equation}
 Y_i(E_\alpha, E_n) = \eta_i\sigma_i(E_\alpha,E_n)dx,
\end{equation}
where $\sigma_i(E_\alpha,E_n)$ is the cross section for this particular interaction. If the material has a total density $\rho$, we define the mass stopping power as $S(E)=-\frac{1}{\rho}\frac{dE}{dx}$. Performing a change in variables and integrating over $E_\alpha$ as the \alp\ particle slows down gives 
\begin{gather}
 \begin{aligned}
  Y^\alpha_i(E_n) &=& \frac{\eta_i}{\rho}\int_0^{E_\alpha}\frac{\sigma_i(E_\alpha^\prime,E_n)}{S(E_\alpha^\prime)}dE_\alpha^\prime \\
                  &=& \frac{N_AC_i}{A_i}\int_0^{E_\alpha}\frac{\sigma_i(E_\alpha^\prime,E_n)}{S(E_\alpha^\prime)}dE_\alpha^\prime, \label{eq:yieldi}
 \end{aligned}
\end{gather}
where $Y^\alpha_i(E_n)$ denotes the thick-target yield of neutrons of energy $E_n$ from a given \alp\ particle (of \emph{initial} energy $E_\alpha$) in the decay chain, $N_A$ is Avogadro's number, $C_i$ is the mass fraction of isotope $i$ in the material, and $A_i$ is the mass number of the target isotope. The total yield for a material can then be found by summing over the yield of each of the target isotopes
\begin{equation}
 Y^\alpha(E_n) = \sum_iY_i^\alpha(E_n).
 \label{eq:yield_mat}
\end{equation}

If we wish to determine the neutron yield of a decay chain consisting of several \alp\ particles, we define $P_\alpha$ to be the probability of an \alp\ particle appearing in a decay of the decay chain, based on the branching ratio for the parent source isotope being produced and the branching ratio for the parent source isotope to decay to an \alp\ particle of this $E_\alpha$. The total yield of neutrons of energy $E_n$ is then given by
\begin{equation}
 Y(E_n) = \sum_\alpha P_\alpha Y^\alpha(E_n).
 \label{eq:yield_total}
\end{equation}
The total number of neutrons produced at any energy is the integral of $Y(E_n)$ over the entire neutron energy spectrum.

The output of \talys\ is the \alphan\ total cross section for an \alp\ particle of specified energy reacting with the specific target nucleus, the individual cross sections for each excited state that the daughter nucleus may occupy after the reaction, cross sections for each \gr\ that may be produced in this reaction, and the energy spectrum of outgoing neutrons. The last quantity is determined by energy and momentum conservation for each daughter nucleus energy level and the corresponding cross section.

\talys\ performs all of its calculations at the specified \alp\ energy; it does not simulate the \alp\ particle slowing down. The output of \talys\ is thus the $\sigma_i$ terms in Equation~\ref{eq:yieldi}. 
It is therefore necessary to integrate over the track of the \alp\ particle as it slows down. This treatment differs from that presented in~\cite{mei_evaluation_2009}, which uses the output of \talys\ directly as the integral, resulting in neutron spectra that predict a higher rate of neutrons at higher energies compared to \nucbot\ and \sources. It also introduces uncertainties in the total neutron yield calculation, though we have not observed a consistent trend compared to \nucbot.

We have compiled a library of \alphan\ reaction cross sections and neutron spectra generated by \talys\ for most naturally occurring isotopes for \alp\ particle energies ranging from 0--10\,MeV in 10\,keV increments. This range is the energy range of \alp\ decays in the naturally occurring uranium and thorium decay chains and is therefore the most relevant to computing \alphan\ neutron background rates.

In order to calculate \alphan\ yields for an arbitrary material, \nucbot\ allows the user to specify a material composition. The user may do so either by giving each isotope and its mass fraction in the material, or by specifying the elemental composition of the material and the mass fraction of each element. In the latter case, the natural abundance of each isotope is looked up from a table published in~\cite{de_bievre_table_1993}, and these abundances are used to determine the isotopic composition of the material. The mass fractions and mass numbers of each isotope are used as $C_i$ and $A_i$, respectively, in Equation~\ref{eq:yieldi}.

The list of \alp\ particle energies and relative intensities can be specified in one of two ways. The user may directly specify these values, or they may provide a list of \alp-emitting isotopes and their relative probabilities of appearing. The latter case is useful for simulating decay chains; the specified probabilities may be chosen as the isotopes' probabilities of appearing in the decay chain. In this case, a list of \alp\ particle energies and branching ratios is looked up for each isotope from the ENSDF database~\cite{tuli_evaluated_1996}. To speed up future computations, ENSDF data files are saved into a local library, so they only need to be retrieved once. These \alp\ particle energies and probabilities are used to define $E_\alpha$ in Equation~\ref{eq:yieldi}, and $P_\alpha$ in Equation~\ref{eq:yield_total}, respectively.

Total neutron yields can be calculated by integrating over the full neutron energy spectrum or by integrating over the total cross sections calculated by \talys. These two methods typically agree to within 1--5\%, deviating primarily due to uncertainties introduced in the Riemann integration and the finite resolution imposed by the binning of the neutron energy spectrum. The total yields output by \nucbot\ are therefore provided by the integral over the total cross sections.

\nucbot\ calculates the neutron yield and energy spectrum for each \alp\ particle as it slows down in a material and interacts with each isotope present, assuming a homogeneous composition (\emph{i.e.}, that each isotope is uniformly distributed in the material), using Equation~\ref{eq:yieldi}. The mass stopping powers $S(E_\alpha)$ are read from a library generated by \srim\ as the \alp\ particle is integrated over its track as it slows down. This integral is approximated with a Riemann sum. The total yield of all neutrons of energy $E_n$ is then found by summing over \alp energies and isotopes, as described by Equations~\ref{eq:yield_mat} and~\ref{eq:yield_total}.

Other information about the \alphan\ reaction, including excited nuclear states and associated coincident \grs\ are included in the TALYS-generated database in \nucbot.

This code, along with several example neutron energy spectra, are discussed in much detail in~\cite{westerdale_thesis}. Neutron yield calculations for several common detector materials whose compositions are given in Table~\ref{tab:matcomp} are shown in Table~\ref{tab:alphanyields}. The decay chains for which neutron yields were computed in this table are described in Section~\ref{sec:decaychains}. The \ura$_{\text{lower}}$ chain in this table includes contributions from \pbten\ and below; the \pbten\ column lists these contributions separately for cases where equilibrium is broken.

\subsection{Calculations from Measured Yields}\label{subsec:measuredyieldcalcs}
Various groups have published compilations of measured \alphan\ yields. In the present discussion, we will draw from the compilations made by~\cite{Heaton:1989eq,roughton_thick-target_1983} for elements between lithium and iron, and the compilation made by ~\cite{stelson_cross_1964} for heavier metals. The sets of measurements from which we drew each \alphan\ yield are summarized in Table~\ref{tab:datarefs}. For these calculations, we assume the material compositions given in Table~\ref{tab:matcomp}.

\begin{table}[ht]
 \centering
 \caption{References for \alphan\ yield measurements for different elements}
 \begin{tabular}{l|l}\hline\hline
  Reference & Elements \\\hline
  Heaton et al.~\cite{Heaton:1989eq} &  Al, Be, B, C, F, Fe, Li, Mg, N, O, Si, Na\\
  Roughton et al.~\cite{roughton_thick-target_1983} & Mn, Ti\\
  Stelson \& McGowan~\cite{stelson_cross_1964} & Co, Cu, Mo, Ni, Ag, Zn, Zr\\\hline\hline
 \end{tabular}
 \label{tab:datarefs}
\end{table}

As discussed in these compilations, these measurements are difficult to perform. As a result, these measurements report uncertainties in the range of 10--20\%, and different groups measuring the \alphan\ yield of the same isotopes often report yields differing by up to 40\%.

Some of the datasets used in these calculations reported neutron yields for elements, assuming all of their isotopes were present at their natural abundance, while others report neutron yields for individual isotopes. In the later case, we assumed that isotopes were present in their natural abundances, as reported in~\cite{de_bievre_table_1993}, and combined these yields to get the average elemental neutron yield, which we use for this comparison. Nuclear decay data, including \alp\ energies and probabilities, are determined using the Evaluated Nuclear Structure Data Files (ENSDF)~\cite{tuli_evaluated_1996}.

When combining measured or calculated thick-target neutron yields over several isotopes to determine the yield of a composite material, it is important to note that the mass stopping power relevant in determining the \alphan\ yield of a material is that material's stopping power. However, measurements made on individual elements are determined by the stopping power of that element alone. In other words, the denominator of Equation~\ref{eq:yieldi} is $S_i(E)$ for each individual isotope being summed over, while it is $S(E)=\sum_iS_i(E)$ for the composite material. 

To account for this difference, we follow the prescription described in~\cite{Heaton:1989eq}, and rewrite Equation~\ref{eq:yieldi} as
\begin{gather}
 \begin{aligned}
 Y^\alpha_i &=&       \frac{N_AC_i}{A_i}\int_0^{E_\alpha}\frac{S_i(E_\alpha^\prime)}{S(E_\alpha^\prime)}\frac{\sigma_i(E_\alpha^\prime,E_n)}{S_i(E_\alpha^\prime)}dE_\alpha^\prime \\
            &\approx& \frac{N_AC_i}{A_i}\frac{S_i(E_\alpha^\prime)}{S(E_\alpha^\prime)}\int_0^{E_\alpha}\frac{\sigma_i(E_\alpha^\prime,E_n)}{S_i(E_\alpha^\prime)}dE_\alpha^\prime \\
            &=& C_i\frac{S_i(E_\alpha^\prime)}{S(E_\alpha^\prime)} \widetilde{Y}^\alpha_i,\hspace{1.25in}
 \end{aligned}
\end{gather}
where $\widetilde{Y}_i^\alpha$ is the measured neutron yield for isotope $i$. The approximation in the second line relies on the assumption that $S_i(E_\alpha^\prime)/S(E_\alpha^\prime)$ is approximately constant as the \alp\ particle slows down---that is to say that the stopping power of each isotope in the material has approximately the same functional form and they differ only by a constant factor~\cite{Heaton:1989eq}.

For this calculation, we assume Bragg's rule for the additivity of stopping powers for the compounds discussed here, which we expect to be a safe assumption~\cite{Heaton:1989eq}. The validity of Bragg's rule is discussed in~\cite{thwaites_braggs_1983}. Heaton et al. estimate that the assumption that $S_i(E_\alpha^\prime)/S(E_\alpha^\prime)$ is approximately constant introduces an uncertainty $\sim5\%$ for $E_\alpha$ in the range of 3--10\,MeV, where \alphan\ neutrons are most likely to be produced. However, the uncertainty is likely higher for lighter nuclei such as carbon and beryllium.

We use \srim\ to calculate the stopping power of each individual element in the material. Heaton et al. estimate a $\sim5\%$ uncertainty in the final neutron yield calculations obtained this way due to uncertainties in \srim.

\begin{table}[ht]
 \centering
 \caption{Isotopes missing from the \sources\ and data compilation calculations presented here. Where only a chemical symbol is given, all isotopes of that element are missing. Li data was only available in \sources\ at low energies, and was included in the \pbten\ and \ura$_{\text{upper}}$ chains.}
 \setlength\extrarowheight{4pt} 
 \footnotesize
 \begin{tabular}{ll|l}\hline\hline
  \multirow{2}{*}{Argon}              & Data        & Ar                                                     \\
                                      & \sources    & $^{36}$Ar, $^{38}$Ar                                   \\\hline
  \multirow{1}{*}{Be-Cu Alloy}        & Data        & $^{61}$Ni, $^{64}$Ni                                   \\\hline
  \multirow{2}{*}{Borosilicate Glass} & Data        & Ba                                                     \\
                                      & \sources    & Ba                                                     \\\hline
  \multirow{2}{*}{Brass}              & Data        & $^{67}$Zn                                              \\
                                      & \sources    & Zn                                                     \\\hline
  \multirow{3}{*}{Kovar}              & Data        & Cr, $^{47}$Ti, $^{61}$Ni, $^{64}$Ni                    \\
                                      & \sources    & Zr, Mo                                                 \\\hline
  \multirow{2}{*}{Solder}             & Data        & Sn                                                     \\
                                      & \sources    & Ag, Sn                                                 \\\hline
  \multirow{2}{*}{Stainless Steel}    & Data        & Cr, P, S, $^{61}$Ni, $^{64}$Ni                         \\
				      & \sources    & S,                                                     \\\hline
  \multirow{1}{*}{Titanium}           & Data        & $^{47}$Ti                                              \\\hline
  \multirow{2}{*}{Xenon}              & Data        & Xe                                                     \\
                                      & \sources    & Xe                                                     \\\hline\hline
 \end{tabular}
 \label{tab:missingisos}
\end{table}

The results of these calculations for several materials common in low-background experiments are summarized in Table~\ref{tab:alphanyields}, where they are compared to \nucbot\ calculations. While these results are derived from measured yields, they are limited by the availability of data. Table~\ref{tab:missingisos} summarizes the isotopes for which we were lacking data for each material whose calculations are presented here. Missing isotopes were omitted from these calculations. These calculations are limited by experimental uncertainties and the uncertainties related to stopping powers discussed above. Additionally, these calculations do not provide neutron spectra, which are necessary for simulating and understanding the neutron backgrounds.
\subsection{Calculations using \sources}\label{subsec:sourcescalcs}

In addition to comparing \nucbot\ neutron yields to the aforementioned data compilations, we used \sources, developed by Los Alamos National Laboratory, to calculate \alphan\ yields of various materials. Validation of \sources\ and the underlying code is discussed at length in~\cite{shores_data_2001}. Calculations presented here use the material compositions given in Table~\ref{tab:matcomp}.

In summary, \sources\ allows the user to specify a material composition and a set of \alp-emitting isotopes. 
It uses \alphan\ cross section and product branching ratio data libraries as well as stopping power calculations to simulate the \alphan\ reaction in the target material. 

By default, \sources\ only contains \alphan\ cross sections for \alp\ energies below 6.5\,MeV, and data for many important isotopes is missing entirely. However, since the input data can be modified by the user, \alphan\ cross sections may be added to extend the libraries. For the comparisons presented here, \sources\ was extended using cross section calculations and measurements from the Japanese Evaluated Nuclear Data Library (\jendl)~\cite{shibata_jendl_2011}, the \empire\ nuclear reaction code~\cite{herman2005empire}, and the work by Stelson \& McGowan~\cite{stelson_cross_1964}. It should be noted that much of the data in JENDL come from the same measurements used in the calculations discussed in Section~\ref{subsec:measuredyieldcalcs}; the \sources\ and data calculations are therefore not completely independent. Table~\ref{tab:sourcesrefs} shows which reference was used for each element in these calculations.

\begin{table}[ht]
 \centering
 \caption{References for \alphan\ yield measurements for different elements, as added to the \sources\ libraries.}
 \begin{tabular}{l|l}\hline\hline
  Reference                                    & Elements                          \\\hline
  \empire~\cite{herman2005empire}              & Mg, P, Ar, Ti, Mn, Fe, Ni         \\
  \jendl~\cite{shibata_jendl_2011}        & Li, Be, B, C, N, O, F, Na, Al, Si \\
  Stelson \& McGowan~\cite{stelson_cross_1964} & Co, Cu                            \\\hline\hline
 \end{tabular} 
 \label{tab:sourcesrefs}
\end{table}

Despite these additions, data for some isotopes was still missing. These isotopes were omitted from the \sources\ yield calculations presented here, and are summarized in Table~\ref{tab:missingisos}.
The results of these \sources\ computations are summarized in Table~\ref{tab:alphanyields}. 

\begin{figure*}[t]
 \includegraphics[width=\linewidth]{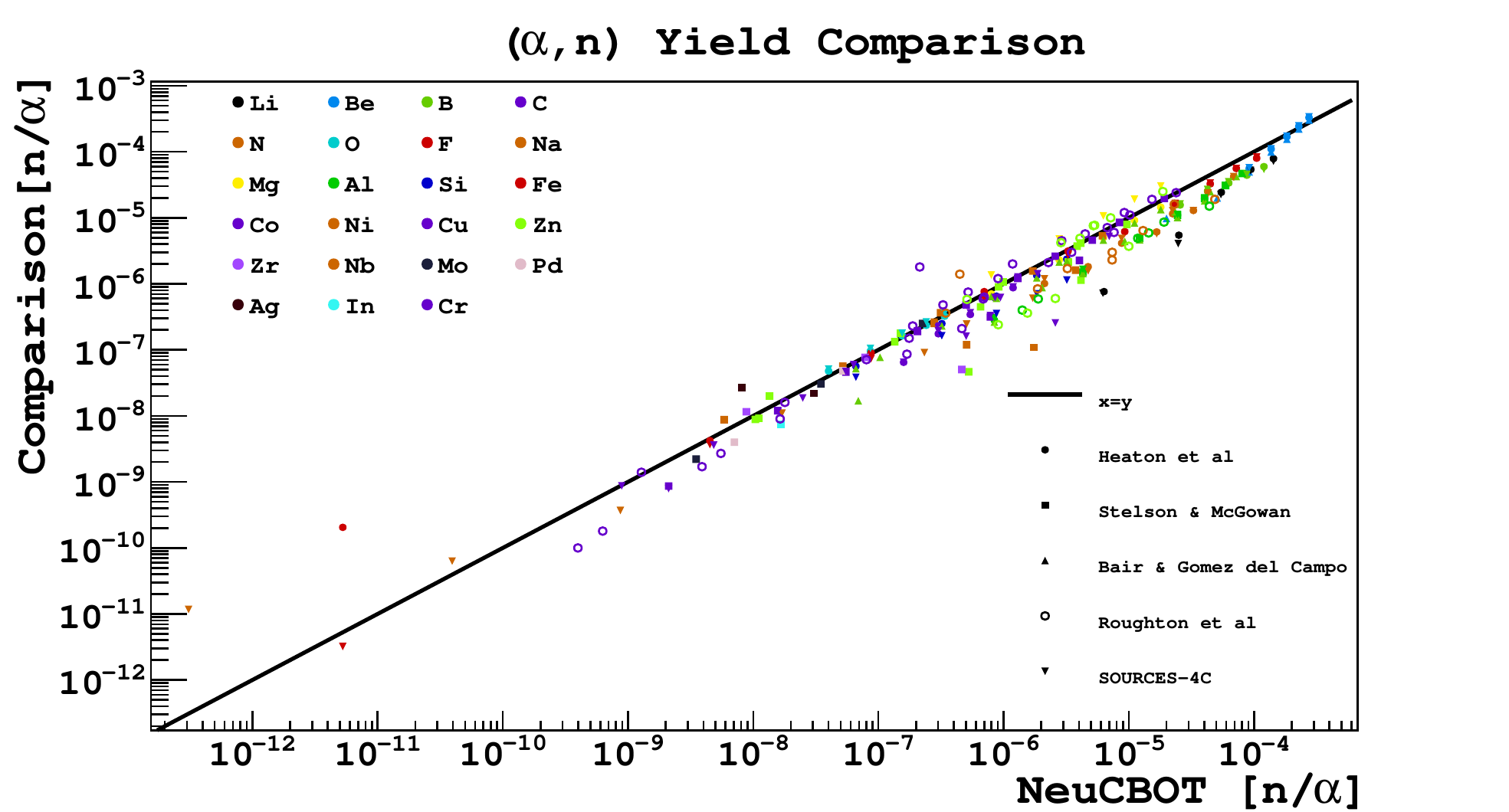}
 \caption{A scatter plot comparing the \alphan\ yields calculated by \nucbot\ to various benchmark measurements and calculations, for \alp\ exposures ranging in energy from 3--11\,MeV. The straight black line is $x=y$, showing where this calculator agrees with a benchmark perfectly. Solid dots represent measurements presented in~\cite{Heaton:1989eq}, squares in~\cite{stelson_cross_1964}, triangles pointing up in~\cite{bair_neutron_1979}, and hollow circles in~\cite{roughton_thick-target_1983}; triangles pointing down are calculations done by \sources. References that reported yields per isotope were compared to \nucbot\ calculations for that isotope; references that reported yields per element were compared to \nucbot\ assuming natural abundances described in the text; \sources\ comparisons were done for elements with these natural abundances. Different colors correspond to measurements done on different target elements; progressions of points the same shape and color represent yields varying over energy.}
 \label{fig:anyield_scatterplot}
\end{figure*}

\subsection{\nucbot\ Validation and Comparisons}
%

\begin{figure}[t]
 \centering
 \includegraphics[width=\linewidth]{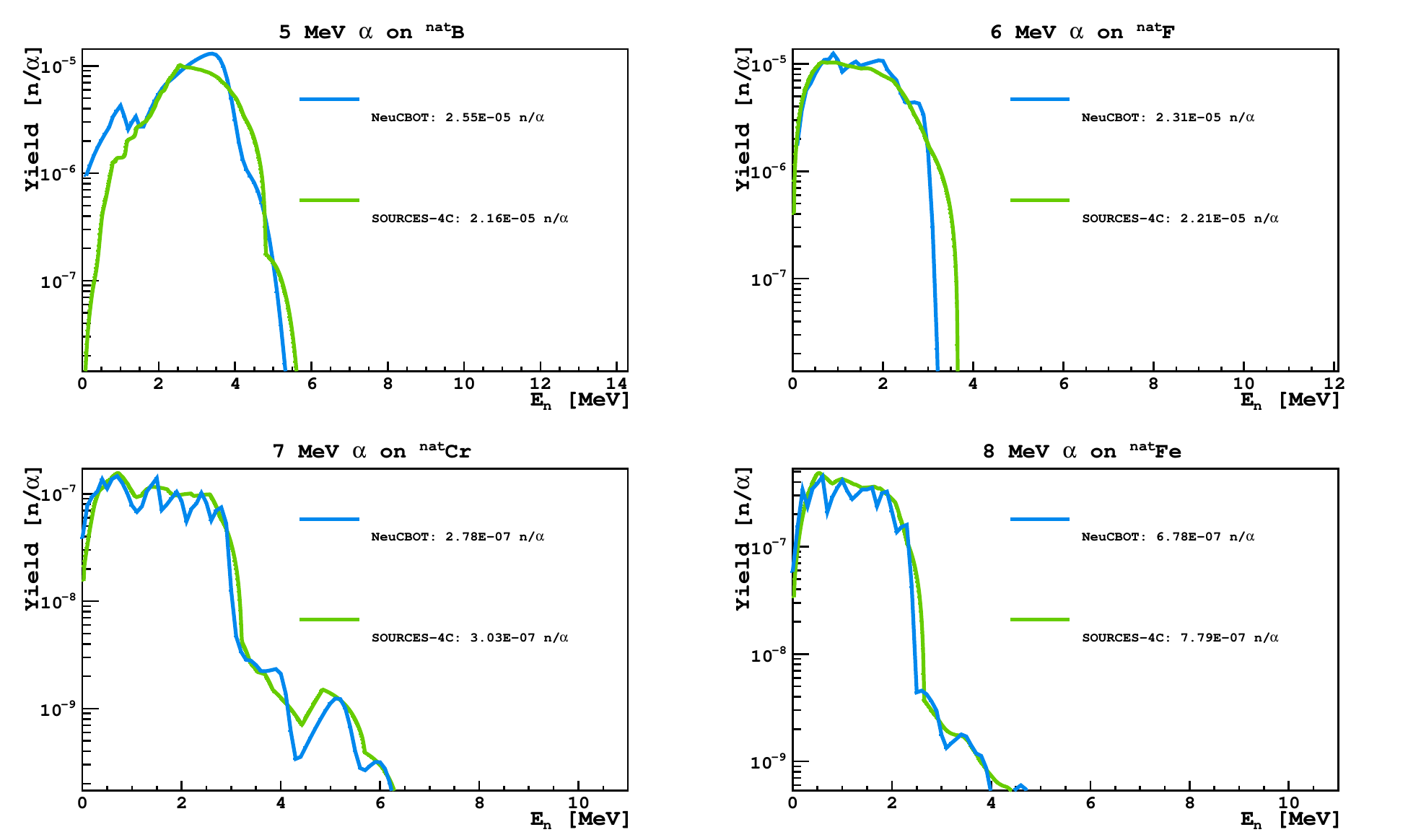}
 \caption{The \alphan\ neutron energy spectrum for 5, 6, 7, and 8\,MeV \alp\ particles stopping in pure boron, fluorine, chromium, and iron targets at their natural isotopic abundances, computed by \nucbot\ (blue) and by \sources\ (green).}
 \label{fig:ancomp_fspec}
\end{figure}

Figure~\ref{fig:anyield_scatterplot} shows a scatter plot comparing various \alphan\ yield measurements and \sources\ calculations to \nucbot\ calculations. 
Qualitatively, this figure shows a strong correlation between data and \nucbot\ calculations over several orders of magnitude. We see that \nucbot\ tends to systematically predict higher yields. An overall evaluation of how \nucbot\ calculations tend to compare with the references show that \nucbot\ yield calculations tend to be systematically higher by $\sim27\%$, with a $\sim35\%$ root mean square on this deviation.



We have found the the outgoing neutron energy spectra predicted by \sources\ and \nucbot\ agree closely, with those generated by \nucbot\ predicting slightly more structure than is seen in the spectra predicted by \sources. A comparison between the spectra predicted for boron, fluorine, chromium, and iron can be seen in Figure~\ref{fig:ancomp_fspec}. More examples are provided in~\cite{westerdale_thesis}.

\section{Conclusions}


We introduce \nucbot, a new tool presented here that calculates \alphan\ yields in arbitrary materials. We have used \nucbot\ to calculate \alphan\ yields in many materials commonly used in low-background experiments due to \ura,\urafive, and \tho\ contamination. We have compared these calculations to calculations made using measured yields and those performed by \sources. While we found that \nucbot\ tends to predict yields systematically higher than these other two tools by $\sim30\%$, it can provide \alphan\ background rate estimates and spectra without being constrained by the availability of measurements.

For completeness, we also include a discussion of other radiogenic production rates due to these contaminants.

This code is available for general use. Documentation and the code can be downloaded at\\ \website.

\section{Acknowledgements}
We would like to thank Pablo Mosteiro and Eric Vazquez Jauregui for help with the \sources\ calculations. We would also like to thank Chris Stanford, Hao Qian, and Guangyong Koh for testing the code and providing valuable feedback.

\begin{table*}[htb]
 \centering
 \caption{\alphan\ yields determined by \nucbot, compared data compilations and \sources. Entries with a dash were not possible to calculate due to lacking data. The \ura$_{\text{lower}}$ chain \emph{includes} equilibrium contributions from \pbten\ and below. $^*$Indicates calculation was done with one or more isotope neglected from the computation due to lacking data.}
 \footnotesize
 \begin{tabular}{ll|ccccc}\hline\hline
                                                &             & \tho\      & \urafive   & \ura$_{\text{upper}}$ & \ura$_{\text{lower}}$ & \pbten     \\
                                                &             & \multicolumn{5}{c}{n/s/Bq}                                                           \\\hline
  \multirow{3}{*}{Acrylic}                      & \nucbot     & 1.33\E{-6} & 1.42\E{-6} & 2.19\E{-7}            & 9.72\E{-7}            & 1.16\E{-7} \\
                                                & Data        & 9.11\E{-7} & 9.28\E{-7} & 1.07\E{-7}            & 6.32\E{-7}            & 6.46\E{-8} \\
                                                & \sources    & 2.42\E{-6} & 2.70\E{-6} & 3.41\E{-7}            & 1.83\E{-6}            & 2.08\E{-7} \\\hline   
  \multirow{3}{*}{Alumina}                      & \nucbot     & 2.21\E{-5} & 2.01\E{-5} & 5.14\E{-7}            & 1.38\E{-5}            & 7.35\E{-7} \\
                                                & Data        & 8.54\E{-6} & 8.54\E{-6} & 2.11\E{-7}            & 5.96\E{-6}            & 2.82\E{-7} \\
                                                & \sources    & 9.87\E{-6} & 8.55\E{-6} & 2.24\E{-7}            & 6.17\E{-6}            & 2.99\E{-7} \\\hline   
  \multirow{3}{*}{Aluminum}                     & \nucbot     & 4.53\E{-5} & 4.12\E{-5} & 1.00\E{-6}            & 2.83\E{-5}            & 1.49\E{-6} \\
                                                & Data        & 1.99\E{-5} & 1.19\E{-5} & 3.37\E{-7}            & 1.19\E{-5}            & 5.26\E{-7} \\
                                                & \sources    & 1.95\E{-5} & 1.67\E{-5} & 3.49\E{-7}            & 1.21\E{-5}            & 5.46\E{-7} \\\hline   
  \multirow{3}{*}{Argon}                        & \nucbot     & 2.64\E{-5} & 1.72\E{-5} & 8.82\E{-8}            & 1.41\E{-5}            & 2.26\E{-7} \\
                                                & Data$^*$    & ---        & ---        & ---                   & ---                   & ---        \\
                                                & \sources$^*$& 2.21\E{-5} & 1.48\E{-5} & 1.77\E{-7}            & 1.18\E{-5}            & 2.31\E{-7} \\\hline   
  \multirow{3}{*}{Be-Cu Alloy}                  & \nucbot     & 6.22\E{-6} & 6.72\E{-6} & 1.58\E{-6}            & 4.66\E{-6}            & 7.59\E{-7} \\
                                                & Data$^*$    & 5.01\E{-6} & 5.26\E{-6} & 8.29\E{-7}            & 3.61\E{-6}            & 4.86\E{-7} \\
                                                & \sources    & 4.86\E{-6} & 5.30\E{-6} & 8.41\E{-7}            & 3.66\E{-6}            & 4.98\E{-7} \\\hline   
  \multirow{3}{*}{Borosilicate Glass}           & \nucbot     & 2.43\E{-5} & 2.56\E{-5} & 3.93\E{-6}            & 1.76\E{-5}            & 2.25\E{-6} \\
                                                & Data$^*$    & 1.33\E{-5} & 1.41\E{-5} & 2.25\E{-6}            & 9.79\E{-6}            & 1.32\E{-6} \\
                                                & \sources$^*$& 1.37\E{-5} & 1.45\E{-5} & 2.38\E{-6}            & 1.08\E{-5}            & 1.46\E{-6} \\\hline   
  \multirow{3}{*}{Brass}                        & \nucbot     & 3.06\E{-7} & 1.42\E{-8} & 6.52\E{-14}           & 2.58\E{-8}            & 5.72\E{-13}\\
                                                & Data$^*$    & 1.81\E{-7} & 1.19\E{-8} & 0                     & 2.31\E{-8}            & 0          \\
                                                & \sources$^*$& 1.59\E{-7} & 9.44\E{-9} & 0                     & 1.82\E{-8}            & 0          \\\hline   
  \multirow{3}{*}{Cirlex}                       & \nucbot     & 3.09\E{-6} & 2.57\E{-6} & 2.66\E{-7}            & 2.01\E{-6}            & 1.39\E{-7} \\
                                                & Data        & 1.64\E{-6} & 1.41\E{-6} & 1.20\E{-7}            & 1.04\E{-6}            & 7.26\E{-8} \\
                                                & \sources    & 1.61\E{-6} & 1.43\E{-6} & 1.22\E{-7}            & 1.08\E{-6}            & 7.49\E{-8} \\\hline   
  \multirow{3}{*}{Copper}                       & \nucbot     & 3.86\E{-7} & 1.71\E{-8} & 0                     & 3.17\E{-8}            & 0          \\
                                                & Data        & 2.13\E{-7} & 1.47\E{-8} & 0                     & 2.88\E{-8}            & 0          \\
                                                & \sources    & 2.53\E{-7} & 1.50\E{-8} & 0                     & 2.90\E{-8}            & 0          \\\hline   
  \multirow{3}{*}{Fused Silica}                 & \nucbot     & 1.81\E{-6} & 1.64\E{-6} & 7.58\E{-8}            & 1.15\E{-6}            & 7.91\E{-8} \\
                                                & Data        & 1.47\E{-6} & 1.37\E{-6} & 8.27\E{-8}            & 9.41\E{-7}            & 7.23\E{-8} \\
                                                & \sources    & 1.47\E{-6} & 1.44\E{-6} & 8.59\E{-8}            & 9.95\E{-7}            & 7.41\E{-8} \\\hline   
  \multirow{3}{*}{Kovar}                        & \nucbot     & 1.22\E{-6} & 2.81\E{-7} & 3.22\E{-9}            & 3.29\E{-7}            & 4.05\E{-9} \\
                                                & Data$^*$    & 1.14\E{-6} & 2.62\E{-7} & 2.19\E{-9}            & 3.31\E{-7}            & 2.56\E{-9} \\
                                                & \sources$^*$& 9.24\E{-7} & 3.59\E{-7} & 1.78\E{-8}            & 3.36\E{-7}            & 1.25\E{-8} \\\hline   
  \multirow{1}{*}{NUV-HD}                       & \nucbot     & 3.51\E{-6} & 3.07\E{-6} & 7.77\E{-8}            & 2.16\E{-6}            & 1.19\E{-7} \\
  Silicon                                       & Data        & 2.63\E{-6} & 2.31\E{-6} & 6.45\E{-8}            & 1.59\E{-6}            & 9.24\E{-8} \\
  Photomultipliers                              & \sources    & 2.58\E{-6} & 2.38\E{-6} & 6.71\E{-8}            & 1.66\E{-6}            & 9.18\E{-8} \\\hline   
  \multirow{3}{*}{Polyethylene Terephthalate}   & \nucbot     & 1.48\E{-6} & 1.58\E{-6} & 2.45\E{-7}            & 1.08\E{-6}            & 1.30\E{-7} \\
                                                & Data        & 1.01\E{-6} & 1.03\E{-6} & 1.19\E{-7}            & 7.01\E{-7}            & 7.18\E{-8} \\
                                                & \sources    & 1.03\E{-6} & 1.11\E{-6} & 1.22\E{-7}            & 7.56\E{-7}            & 7.49\E{-8} \\\hline   
  \multirow{3}{*}{PTFE}                         & \nucbot     & 1.27\E{-4} & 1.31\E{-4} & 1.16\E{-5}            & 8.85\E{-5}            & 9.37\E{-6} \\
                                                & Data        & 9.26\E{-5} & 9.56\E{-5} & 8.15\E{-6}            & 6.46\E{-5}            & 6.33\E{-6} \\
                                                & \sources    & 8.94\E{-5} & 9.53\E{-5} & 7.19\E{-6}            & 6.40\E{-5}            & 6.08\E{-6} \\\hline   
  \multirow{3}{*}{Solder}                       & \nucbot     & 2.53\E{-9} & 1.11\E{-10}& 0                     & 2.06\E{-10}           & 0          \\
                                                & Data$^*$    & 1.37\E{-9} & 9.53\E{-11}& 0                     & 1.87\E{-10}           & 0          \\
                                                & \sources$^*$& 1.64\E{-9} & 9.68\E{-11}& 0                     & 1.87\E{-10}           & 0          \\\hline   
  \multirow{3}{*}{Stainless Steel}              & \nucbot     & 1.96\E{-6} & 4.42\E{-7} & 1.31\E{-9}            & 5.52\E{-7}            & 2.14\E{-9} \\
                                                & Data$^*$    & 1.25\E{-6} & 2.85\E{-7} & 1.04\E{-9}            & 3.73\E{-7}            & 1.43\E{-9} \\
                                                & \sources$^*$& 1.57\E{-6} & 3.59\E{-7} & 1.14\E{-9}            & 4.38\E{-7}            & 1.72\E{-9} \\\hline   
  \multirow{3}{*}{Titanium}                     & \nucbot     & 7.34\E{-6} & 2.58\E{-6} & 2.89\E{-9}            & 2.81\E{-6}            & 1.17\E{-8} \\
                                                & Data$^*$    & 6.30\E{-6} & 2.75\E{-6} & 0                     & 2.60\E{-6}            & 2.13\E{-8} \\
                                                & \sources    & 5.41\E{-6} & 1.93\E{-6} & 2.19\E{-9}            & 2.05\E{-6}            & 8.56\E{-9} \\\hline   
  \multirow{3}{*}{Viton}                        & \nucbot     & 1.15\E{-4} & 1.19\E{-4} & 1.06\E{-5}            & 8.07\E{-5}            & 8.53\E{-6} \\
                                                & Data        & 8.44\E{-5} & 8.72\E{-5} & 7.43\E{-6}            & 5.89\E{-5}            & 5.77\E{-6} \\
                                                & \sources    & 8.11\E{-5} & 8.65\E{-5} & 6.52\E{-6}            & 5.81\E{-5}            & 5.51\E{-6} \\\hline   
  \multirow{3}{*}{Xenon}                        & \nucbot     & 6.15\E{-12}& 1.24\E{-14}& 0                     & 1.25\E{-13}           & 0          \\
			                        & Data$^*$    & ---        & ---        & ---                   & ---                   & ---        \\
			                        & \sources$^*$& ---        & ---        & ---                   & ---                   & ---        \\\hline
 \end{tabular}
 \label{tab:alphanyields}
\end{table*}

\bibliographystyle{nucbot}
\bibliography{biblio}

\begin{thebibliography}{34}
\expandafter\ifx\csname natexlab\endcsname\relax\def\natexlab#1{#1}\fi
\expandafter\ifx\csname bibnamefont\endcsname\relax
  \def\bibnamefont#1{#1}\fi
\expandafter\ifx\csname bibfnamefont\endcsname\relax
  \def\bibfnamefont#1{#1}\fi
\expandafter\ifx\csname citenamefont\endcsname\relax
  \def\citenamefont#1{#1}\fi
\expandafter\ifx\csname url\endcsname\relax
  \def\url#1{\texttt{#1}}\fi
\expandafter\ifx\csname urlprefix\endcsname\relax\def\urlprefix{URL }\fi
\providecommand{\bibinfo}[2]{#2}
\providecommand{\eprint}[2][]{\url{#2}}

\bibitem{exo-200_collaboration_investigation_2015}
\bibinfo{author}{\bibnamefont{{EXO-200 Collaboration}}}, \bibnamefont{et~al.},
  \emph{\bibinfo{title}{Investigation of radioactivity-induced backgrounds in
  {EXO}-200}},
  \href{http://dx.doi.org/10.1103/PhysRevC.92.015503}{\bibinfo{journal}{Physical
  Review C}, \textbf{\bibinfo{volume}{92}},
  \bibinfo{pages}{015503}\bibinfo{year}{ (\bibinfo{year}{2015})}}.

\bibitem{ackermann_gerda_2013}
\bibinfo{author}{\bibfnamefont{K.-H.} \bibnamefont{Ackermann}},
  \bibnamefont{et~al.}, \emph{\bibinfo{title}{The {Gerda} experiment for the
  search of $0\nu\beta\beta$ decay in $^{76}$ge}},
  \href{http://dx.doi.org/10.1140/epjc/s10052-013-2330-0}{\bibinfo{journal}{The
  European Physical Journal C}, \textbf{\bibinfo{volume}{73}},
  \bibinfo{pages}{2330}\bibinfo{year}{ (\bibinfo{year}{2013})}}, ISSN
  \bibinfo{issn}{1434-6044, 1434-6052}.

\bibitem{andringa_current_2016}
\bibinfo{author}{\bibfnamefont{S.}~\bibnamefont{Andringa}},
  \bibnamefont{et~al.}, \emph{\bibinfo{title}{Current {Status} and {Future}
  {Prospects} of the {SNO}+ {Experiment}}},
  \href{http://dx.doi.org/10.1155/2016/6194250}{\bibinfo{journal}{Advances in
  High Energy Physics}, \textbf{\bibinfo{volume}{2016}},
  \bibinfo{pages}{e6194250}\bibinfo{year}{ (\bibinfo{year}{2016})}}, ISSN
  \bibinfo{issn}{1687-7357}.

\bibitem{DarkSide_UAr_2015}
\bibinfo{author}{\bibfnamefont{P.}~\bibnamefont{Agnes}}, \bibnamefont{et~al.}
  (\bibinfo{collaboration}{The DarkSide Collaboration}),
  \emph{\bibinfo{title}{Results from the first use of low radioactivity argon
  in a dark matter search}},
  \href{http://dx.doi.org/10.1103/PhysRevD.93.081101}{\bibinfo{journal}{Phys.
  Rev. D}, \textbf{\bibinfo{volume}{93}},
  \bibinfo{pages}{081101}\bibinfo{year}{ (\bibinfo{year}{2016})}}.

\bibitem{Boulay_deap-3600_2012}
\bibinfo{author}{\bibfnamefont{M.~G.} \bibnamefont{Boulay}}
  (\bibinfo{collaboration}{{The} {DEAP} {Collaboration}}),
  \emph{\bibinfo{title}{{DEAP}-3600 dark matter search at {SNOLAB}}},
  \href{http://dx.doi.org/10.1088/1742-6596/375/1/012027}{\bibinfo{journal}{J.
  Phys. Conf. Ser.}, \textbf{\bibinfo{volume}{375}},
  \bibinfo{pages}{012027}\bibinfo{year}{ (\bibinfo{year}{2012})}}, ISSN
  \bibinfo{issn}{1742-6596}.

\bibitem{aprile_lowering_2015}
\bibinfo{author}{\bibfnamefont{E.}~\bibnamefont{Aprile}}, \bibnamefont{et~al.},
  \emph{\bibinfo{title}{Lowering the radioactivity of the photomultiplier tubes
  for the {XENON}1t dark matter experiment}},
  \href{http://dx.doi.org/10.1140/epjc/s10052-015-3657-5}{\bibinfo{journal}{The
  European Physical Journal C}, \textbf{\bibinfo{volume}{75}},
  \bibinfo{pages}{1}\bibinfo{year}{ (\bibinfo{year}{2015})}}, ISSN
  \bibinfo{issn}{1434-6044, 1434-6052}.

\bibitem{Malling_after_2011}
\bibinfo{author}{\bibfnamefont{D.~C.} \bibnamefont{Malling}},
  \bibnamefont{et~al.}, \emph{\bibinfo{title}{After {LUX}: {The} {LZ}
  {Program}}},
  \href{http://arxiv.org/abs/1110.0103}{\bibinfo{journal}{arXiv}:1110.0103\bibinfo{year}{
  (\bibinfo{year}{2011})}}.

\bibitem{CDMS2015_Ge}
\bibinfo{author}{\bibfnamefont{R.}~\bibnamefont{Agnese}}, \bibnamefont{et~al.}
  (\bibinfo{collaboration}{The {SuperCDMS} Collaboration}),
  \emph{\bibinfo{title}{Improved {WIMP}-search reach of the {CDMS II} germanium
  data}},
  \href{http://dx.doi.org/10.1103/PhysRevD.92.072003}{\bibinfo{journal}{Phys.
  Rev. D}, \textbf{\bibinfo{volume}{92}},
  \bibinfo{pages}{072003}\bibinfo{year}{ (\bibinfo{year}{2015})}}.

\bibitem{agnes_veto_2016}
\bibinfo{author}{\bibfnamefont{P.}~\bibnamefont{Agnes}}, \bibnamefont{et~al.},
  \emph{\bibinfo{title}{The veto system of the {DarkSide}-50 experiment}},
  \href{http://dx.doi.org/10.1088/1748-0221/11/03/P03016}{\bibinfo{journal}{J.
  Instrum.}, \textbf{\bibinfo{volume}{11}},
  \bibinfo{pages}{P03016}\bibinfo{year}{ (\bibinfo{year}{2016})}}, ISSN
  \bibinfo{issn}{1748-0221}.

\bibitem{empl_fluka_2014}
\bibinfo{author}{\bibfnamefont{A.}~\bibnamefont{Empl}},
  \bibinfo{author}{\bibfnamefont{E.}~\bibnamefont{Hungerford}},
  \bibinfo{author}{\bibfnamefont{R.}~\bibnamefont{Jasim}}, \bibnamefont{and}
  \bibinfo{author}{\bibfnamefont{P.}~\bibnamefont{Mosteiro}},
  \emph{\bibinfo{title}{A {Fluka} study of underground cosmogenic neutron
  production}},
  \href{http://dx.doi.org/10.1088/1475-7516/2014/08/064}{\bibinfo{journal}{J.
  Cosmol. Astropart. Phys.}, \textbf{\bibinfo{volume}{2014}},
  \bibinfo{pages}{064}\bibinfo{year}{ (\bibinfo{year}{2014})}}, ISSN
  \bibinfo{issn}{1475-7516}.

\bibitem{westerdale_prototype_2016}
\bibinfo{author}{\bibfnamefont{S.}~\bibnamefont{Westerdale}},
  \bibinfo{author}{\bibfnamefont{E.}~\bibnamefont{Shields}}, \bibnamefont{and}
  \bibinfo{author}{\bibfnamefont{F.}~\bibnamefont{Calaprice}},
  \emph{\bibinfo{title}{A prototype neutron veto for dark matter detectors}},
  \href{http://dx.doi.org/10.1016/j.astropartphys.2016.01.005}{\bibinfo{journal}{Astropart.
  Phys.}, \textbf{\bibinfo{volume}{79}}, \bibinfo{pages}{10}\bibinfo{year}{
  (\bibinfo{year}{2016})}}, ISSN \bibinfo{issn}{09276505}.

\bibitem{maneschg_measurements_2008}
\bibinfo{author}{\bibfnamefont{W.}~\bibnamefont{Maneschg}},
  \bibnamefont{et~al.}, \emph{\bibinfo{title}{Measurements of extremely low
  radioactivity levels in stainless steel for {GERDA}}},
  \href{http://dx.doi.org/10.1016/j.nima.2008.05.036}{\bibinfo{journal}{Nuclear
  Instruments and Methods in Physics Research Section A: Accelerators,
  Spectrometers, Detectors and Associated Equipment},
  \textbf{\bibinfo{volume}{593}}, \bibinfo{pages}{448}\bibinfo{year}{
  (\bibinfo{year}{2008})}}, ISSN \bibinfo{issn}{0168-9002}.

\bibitem{gdms_evans}
\bibinfo{author}{\bibnamefont{{Evans Analytical Group}}},
  \emph{\href{http://www.eaglabs.com/mc/glow-discharge-mass-spectrometry.html}{\bibinfo{title}{http://www.eaglabs.com/mc/glow-discharge-mass-spectrometry.html}}}.

\bibitem{basunia2014nuclear}
\bibinfo{author}{\bibfnamefont{M.}~\bibnamefont{Shamsuzzoha~Basunia}},
  \emph{\bibinfo{title}{Nuclear {Data} {Sheets} for {A} = 210}},
  \href{http://dx.doi.org/10.1016/j.nds.2014.09.004}{\bibinfo{journal}{Nuclear
  Data Sheets}, \textbf{\bibinfo{volume}{121}},
  \bibinfo{pages}{561}\bibinfo{year}{ (\bibinfo{year}{2014})}}, ISSN
  \bibinfo{issn}{0090-3752}.

\bibitem{valentine_evaluation_2001}
\bibinfo{author}{\bibfnamefont{T.~E.} \bibnamefont{Valentine}},
  \emph{\bibinfo{title}{Evaluation of prompt fission gamma rays for use in
  simulating nuclear safeguard measurements}},
  \href{http://dx.doi.org/10.1016/S0306-4549(00)00039-6}{\bibinfo{journal}{Ann.
  Nucl. Energy}, \textbf{\bibinfo{volume}{28}},
  \bibinfo{pages}{191}\bibinfo{year}{ (\bibinfo{year}{2001})}}, ISSN
  \bibinfo{issn}{03064549}.

\bibitem{holden1984reevaluation}
\bibinfo{author}{\bibfnamefont{N.~E.} \bibnamefont{Holden}} \bibnamefont{and}
  \bibinfo{author}{\bibfnamefont{M.~S.} \bibnamefont{Zucker}},
  \href{http://www.iaea.org/inis/collection/NCLCollectionStore/_Public/16/052/16052650.pdf}{\emph{\bibinfo{title}{A
  reevaluation of the average prompt neutron emission multiplicity (nubar)
  values from fission of uranium and transuranium nuclides}}},
  \bibinfo{publisher}{National Nuclear Data Center, Brookhaven National
  Laboratory} (\bibinfo{year}{1984}).

\bibitem{de_bievre_table_1993}
\bibinfo{author}{\bibfnamefont{P.}~\bibnamefont{De~Biévre}} \bibnamefont{and}
  \bibinfo{author}{\bibfnamefont{P.~D.~P.} \bibnamefont{Taylor}},
  \emph{\bibinfo{title}{Table of the isotopic compositions of the elements}},
  \href{http://dx.doi.org/10.1016/0168-1176(93)87009-H}{\bibinfo{journal}{Int.
  J. Mass Spectrom.}, \textbf{\bibinfo{volume}{123}},
  \bibinfo{pages}{149}\bibinfo{year}{ (\bibinfo{year}{1993})}}, ISSN
  \bibinfo{issn}{0168-1176}.

\bibitem{materion_protherm_becualloy}
\bibinfo{author}{\bibnamefont{{Materion Brush Performance Alloys}}},
  \emph{\href{http://materion.com/~/media/Files/PDFs/Alloy/DataSheets/MoldMAX/AD0060-0311\%20PROtherm.pdf}{\bibinfo{title}{PROtherm\texttrademark
  Alloy Data Sheet}}}.

\bibitem{berger1998stopping}
\bibinfo{author}{\bibfnamefont{M.~J.} \bibnamefont{Berger}},
  \bibinfo{author}{\bibfnamefont{J.}~\bibnamefont{Coursey}},
  \bibinfo{author}{\bibfnamefont{M.}~\bibnamefont{Zucker}}, \bibnamefont{and}
  \bibinfo{author}{\bibfnamefont{J.}~\bibnamefont{Chang}},
  \href{http://physics.nist.gov/PhysRefData/Star/Text/ESTAR.html}{\emph{\bibinfo{title}{Stopping-power
  and range tables for electrons, protons, and helium ions}}},
  \bibinfo{publisher}{NIST Physics Laboratory} (\bibinfo{year}{1998}).

\bibitem{carpenter_kovar_datasheet}
\bibinfo{author}{\bibnamefont{{Carpenter Technology Corporation}}},
  \emph{\href{http://cartech.ides.com/datasheet.aspx?i=101&e=173&c=TechArt}{\bibinfo{title}{Kovar\textsuperscript{\textregistered}
  Alloy}}}.

\bibitem{ssa_stainless_specs}
\bibinfo{author}{\bibnamefont{{Southwest Stainless \& Alloy}}},
  \emph{\href{http://www.stainlesstubularproducts.com/pdf/alloy_information_specifications.pdf}{\bibinfo{title}{ALLOY
  INFORMATION AND SPECS}}}.

\bibitem{koning2013talys}
\bibinfo{author}{\bibfnamefont{A.}~\bibnamefont{Koning}},
  \bibinfo{author}{\bibfnamefont{S.}~\bibnamefont{Hilaire}}, \bibnamefont{and}
  \bibinfo{author}{\bibfnamefont{S.}~\bibnamefont{Goriely}},
  \emph{\bibinfo{title}{{TALYS-1.6}}},
  \href{http://www.talys.eu/fileadmin/talys/user/docs/talys1.6.pdf}{\bibinfo{journal}{Nuclear
  Reaction Program}, \bibinfo{year}{ (\bibinfo{year}{2013})}}.

\bibitem{tuli_evaluated_1996}
\bibinfo{author}{\bibfnamefont{J.~K.} \bibnamefont{Tuli}},
  \emph{\bibinfo{title}{Evaluated nuclear structure data file}},
  \href{http://dx.doi.org/10.1016/S0168-9002(96)80040-4}{\bibinfo{journal}{Nucl.
  Instrum. Meth. A}, \textbf{\bibinfo{volume}{369}},
  \bibinfo{pages}{506}\bibinfo{year}{ (\bibinfo{year}{1996})}}, ISSN
  \bibinfo{issn}{0168-9002}.

\bibitem{ziegler1985stopping}
\bibinfo{author}{\bibfnamefont{J.~F.} \bibnamefont{Ziegler}} \bibnamefont{and}
  \bibinfo{author}{\bibfnamefont{J.~P.} \bibnamefont{Biersack}},
  \emph{\bibinfo{title}{The stopping and range of ions in matter}},
  \bibinfo{publisher}{Springer} (\bibinfo{year}{1985}).

\bibitem{mei_evaluation_2009}
\bibinfo{author}{\bibfnamefont{D.~M.} \bibnamefont{Mei}},
  \bibinfo{author}{\bibfnamefont{C.}~\bibnamefont{Zhang}}, \bibnamefont{and}
  \bibinfo{author}{\bibfnamefont{A.}~\bibnamefont{Hime}},
  \emph{\bibinfo{title}{Evaluation of induced neutrons as a background for dark
  matter experiments}},
  \href{http://dx.doi.org/10.1016/j.nima.2009.04.032}{\bibinfo{journal}{Nucl.
  Instrum. Meth. A}, \textbf{\bibinfo{volume}{606}},
  \bibinfo{pages}{651}\bibinfo{year}{ (\bibinfo{year}{2009})}}, ISSN
  \bibinfo{issn}{0168-9002}.

\bibitem{westerdale_thesis}
\bibinfo{author}{\bibfnamefont{S.}~\bibnamefont{Westerdale}},
  \href{http://web.archive.org/web/20161216231000/https://www.princeton.edu/physics/graduate-program/theses/Westerdalethesis.pdf}{\textit{A
  Study of Nuclear Recoil Backgrounds in Dark Matter Detectors}}, Ph.D. thesis,
  \bibinfo{school}{Princeton University} (\bibinfo{year}{2016}).

\bibitem{Heaton:1989eq}
\bibinfo{author}{\bibfnamefont{R.}~\bibnamefont{Heaton}},
  \bibinfo{author}{\bibfnamefont{H.}~\bibnamefont{Lee}},
  \bibinfo{author}{\bibfnamefont{P.}~\bibnamefont{Skensved}}, \bibnamefont{and}
  \bibinfo{author}{\bibfnamefont{B.~C.} \bibnamefont{Robertson}},
  \emph{\bibinfo{title}{{Neutron production from thick-target ($\alpha$, n)
  reactions}}},
  \href{http://dx.doi.org/10.1016/0168-9002(89)90579-2}{\bibinfo{journal}{Nucl.
  Inst. Meth. A}, \textbf{\bibinfo{volume}{276}},
  \bibinfo{pages}{529}\bibinfo{year}{ (\bibinfo{year}{1989})}}.

\bibitem{roughton_thick-target_1983}
\bibinfo{author}{\bibfnamefont{N.~A.} \bibnamefont{Roughton}},
  \bibnamefont{et~al.}, \emph{\bibinfo{title}{Thick-target measurements and
  astrophysical thermonuclear reaction rates: $\alpha$-induced reactions}},
  \href{http://dx.doi.org/10.1016/0092-640X(83)90021-9}{\bibinfo{journal}{At.
  Data Nucl. Data Tables}, \textbf{\bibinfo{volume}{28}},
  \bibinfo{pages}{341}\bibinfo{year}{ (\bibinfo{year}{1983})}}, ISSN
  \bibinfo{issn}{0092-640X}.

\bibitem{stelson_cross_1964}
\bibinfo{author}{\bibfnamefont{P.~H.} \bibnamefont{Stelson}} \bibnamefont{and}
  \bibinfo{author}{\bibfnamefont{F.~K.} \bibnamefont{McGowan}},
  \emph{\bibinfo{title}{Cross {Sections} for ($\alpha$,n) {Reactions} for
  {Medium}-{Weight} {Nuclei}}},
  \href{http://dx.doi.org/10.1103/PhysRev.133.B911}{\bibinfo{journal}{Phys.
  Rev.}, \textbf{\bibinfo{volume}{133}}, \bibinfo{pages}{B911}\bibinfo{year}{
  (\bibinfo{year}{1964})}}.

\bibitem{thwaites_braggs_1983}
\bibinfo{author}{\bibfnamefont{D.~I.} \bibnamefont{Thwaites}},
  \emph{\bibinfo{title}{Bragg's {Rule} of {Stopping} {Power} {Additivity}: {A}
  {Compilation} and {Summary} of {Results}}},
  \href{http://dx.doi.org/10.2307/3576096}{\bibinfo{journal}{Radiation
  Research}, \textbf{\bibinfo{volume}{95}}, \bibinfo{pages}{495}\bibinfo{year}{
  (\bibinfo{year}{1983})}}, ISSN \bibinfo{issn}{0033-7587}.

\bibitem{shores_data_2001}
\bibinfo{author}{\bibfnamefont{E.~F.} \bibnamefont{Shores}},
  \emph{\bibinfo{title}{Data updates for the {SOURCES-4A} computer code}},
  \href{http://dx.doi.org/10.1016/S0168-583X(00)00694-7}{\bibinfo{journal}{Nucl.
  Instrum. Meth. B}, \textbf{\bibinfo{volume}{179}},
  \bibinfo{pages}{78}\bibinfo{year}{ (\bibinfo{year}{2001})}}, ISSN
  \bibinfo{issn}{0168-583X}.

\bibitem{shibata_jendl_2011}
\bibinfo{author}{\bibfnamefont{K.}~\bibnamefont{Shibata}},
  \bibnamefont{et~al.}, \emph{\bibinfo{title}{{JENDL}-4.0: {A} {New} {Library}
  for {Nuclear} {Science} and {Engineering}}},
  \href{http://dx.doi.org/10.1080/18811248.2011.9711675}{\bibinfo{journal}{Journal
  of Nuclear Science and Technology}, \textbf{\bibinfo{volume}{48}},
  \bibinfo{pages}{1}\bibinfo{year}{ (\bibinfo{year}{2011})}}, ISSN
  \bibinfo{issn}{0022-3131}.

\bibitem{herman2005empire}
\bibinfo{author}{\bibfnamefont{M.}~\bibnamefont{Herman}}, \bibnamefont{et~al.},
  \emph{\bibinfo{title}{Empire nuclear reaction model code, version 2.19
  (lodi)}},
  \href{https://nucleus.iaea.org/Pages/nuclear-reaction-model-code.aspx}{\bibinfo{journal}{Nuclear
  Data Sheets}, \textbf{\bibinfo{volume}{108}},
  \bibinfo{pages}{2655}\bibinfo{year}{ (\bibinfo{year}{2005})}}.

\bibitem{bair_neutron_1979}
\bibinfo{author}{\bibfnamefont{J.~K.} \bibnamefont{Bair}} \bibnamefont{and}
  \bibinfo{author}{\bibfnamefont{J.}~\bibnamefont{Gomez~del Campo}},
  \emph{\bibinfo{title}{Neutron {Yields} from {Alpha}-{Particle}
  {Bombardment}}},
  \href{http://dx.doi.org/10.13182/NSE71-18}{\bibinfo{journal}{Nucl. Sci.
  Eng.}, \textbf{\bibinfo{volume}{71}}, \bibinfo{pages}{18}\bibinfo{year}{
  (\bibinfo{year}{1979})}}, ISSN \bibinfo{issn}{00295639}.

\end{thebibliography}

\end{document}